\documentclass[smallextended]{svjour3}

\RequirePackage{fix-cm}
\smartqed
\usepackage{graphicx}

\usepackage{amsmath,amsfonts}
\usepackage{algorithmic}
\usepackage{array}
\usepackage[caption=false,font=normalsize,labelfont=sf,textfont=sf]{subfig}
\usepackage{textcomp}
\usepackage{url}
\usepackage{graphicx}
\usepackage{adjustbox}
\usepackage{placeins}
\usepackage{subcaption}
\usepackage[noadjust]{cite}
\usepackage{float}
\usepackage{amsmath}
\usepackage{tipa}
\pdfglyphtounicode{tfm:tipa10/at}{0259}

\usepackage{balance}

\begin{document}

\title{Design and Verification of a Terahertz Bandpass Filter using a Spoof Surface Plasmon Polariton Waveguide with Gapped Unit Cells}

\author{Mohsen Haghighat$^{1}$ \and Ali~Dehghanian$^{1}$  \and
	Levi Smith$^{1,2,*}$
    }

\institute{1. Department of Electrical and Computer Engineering, University of Victoria, Victoria, BC V8P 5C2, Canada.
\\
2. Centre for Advanced Materials and Related Technology (CAMTEC), University of Victoria, 3800 Finnerty Rd, Victoria, BC V8P 5C2, Canada.
\\
* Corresponding author;	\email{levismith@uvic.ca}
}

\date{Received: date / Accepted: date}

\maketitle

\begin{abstract}
This paper presents the experimental verification of a planar guided-wave terahertz (THz) spoof surface plasmon polariton (SSPP) bandpass filter (BPF) using a coplanar stripline (CPS) with internal grooves and periodic gaps. The proposed BPF operates by combining the low-pass behavior from the SSPP's band edge and the high-pass behavior from the gaps that act as series capacitors. The higher and lower cut-off frequencies can be tailored by the appropriate selection of the unit cell geometry. For demonstration, a BPF with a center frequency of approximately 1 THz and a bandwidth of 0.25 THz was designed, fabricated, and experimentally validated. The passband around 1 THz is observed in the measurements, along with the lower and higher cut-off frequencies at approximately 0.91 THz and 1.16 THz, respectively, in agreement with simulation results.

\end{abstract}

\section{Introduction}

Spoof surface plasmon polariton (SSPP) devices operating at microwave and terahertz (THz) frequencies have garnered substantial research interest over the past two decades \cite{Pendry2004_Mimicking,garcia-vidal_surfaces_2005,Review_of_SSPP_Tang_2019,xu_terahertz_2019,unutmaz_terahertz_2019, Haghighat2024_NSRep}. This attention is primarily due to their inherited properties from conventional surface plasmon polaritons (SPPs), including significant field confinement and customizable dispersion characteristics through geometric modifications \cite{shen_conformal_2013,maier_terahertz_2006,maier2007-plasmonics-book, huidobro_pendry_vidal_book_2018}. Originally, SSPP structures were introduced using 3D geometries, such as arrays of holes or grooves on a metal surface, interacting with obliquely incident light or waves \cite{Pendry2004_Mimicking,garcia-vidal_surfaces_2005}. Subsequently, the use of guided waves for SSPP excitation was explored due to their improved integration possibilities \cite{Review_of_SSPP_Tang_2019}. A typical guided-wave SSPP structure comprises a corrugated single conductor and a matching or transition circuit (TC).

Most SSPP research has focused on low-loss SSPP waveguides \cite{unutmaz_terahertz_2019} or their \textit{low-pass} filtering capabilities \cite{guo_spoof_2018}. Few SSPP-based \textit{band-pass} filters (BPF) have been reported beyond microwave and millimeter wave frequencies \cite{guo_novel_BPF_IEEE_access_2018,Wang2019_BPF_SSPP_10GHz_IEEEAccess,Wei2020_BPF_SSPP_IEEE_plasma_sci,Liu-Xu2022_BPF_SSPP_80GHz_TMTT,Feng-Xu2024_BPF_SSPP_30GHz, ren2020leaky, haghighat2026terahertz}. In general, the mentioned BPFs either use cavities, which make on-chip integration challenging, or use a microstrip configuration, which is susceptible to significant radiation losses on standard substrates at THz frequencies  \cite{Levi_Smith_CPS_on_Si3N4_1st}. Furthermore, there is limited research on experimentally validated SSPP-based BPFs at THz frequencies, mainly due to a lack of instrumentation, coupling challenges, and various loss mechanisms \cite{THz_Communication_challanges_IEEE_TTST_2021, Review_of_SSPP_Tang_2019}. In addition to these challenges in the design of SSPP-based BPFs, practical implementation at THz frequencies is complicated by excitation and detection difficulties. One of the popular feedlines for the excitation of SSPPs is coplanar waveguides (CPW) \cite{unutmaz_terahertz_2019}. Integration of CPW with SSPPs presents challenges, including the need for large flaring grounds to efficiently excite SSPPs \cite{unutmaz_investigation_2022}. As an alternative, CPS feedlines have demonstrated their effectiveness at THz frequencies in achieving low-loss SSPP excitation \cite{Haghighat2024_NSRep}. SSPP waveguides that are integrated with CPS can have an internal \cite{Haghighat_Internal_SSPP_24_OE} or external \cite{guo_spoof_2018} array of stubs, depending on the specific application.

BPFs are one of the key components of front-end systems for wireless communication applications, used for band and channel selection and the rejection of unwanted signals \cite{Feng-Xu2024_BPF_SSPP_30GHz, Liu-Xu2022_BPF_SSPP_80GHz_TMTT}. Therefore, the SSPP-based BPF, with the merits of low loss, broadband capability, and a compact structure, has been widely investigated at microwave frequencies in the past decade \cite{Feng-Xu2024_BPF_SSPP_30GHz,guo_novel_BPF_IEEE_access_2018,Liu-Xu2022_BPF_SSPP_80GHz_TMTT,Wei2020_BPF_SSPP_IEEE_plasma_sci,Wang2019_BPF_SSPP_10GHz_IEEEAccess}. Additionally, BPFs are useful components for sensing applications, targeting a specific range of absorption frequencies \cite{Lee_2015_Highly_gluceose_sensing_nano_antennas_nsrep}. Nevertheless, the literature reveals a gap in the design and experimental demonstration of planar (guided-wave) THz SSPP-based band-pass filters, particularly for frequencies beyond 300 GHz, which have potential applications in future short-range THz data communication and sensing applications. 

\begin{figure}
    \centering
    \includegraphics[width=0.4\linewidth]{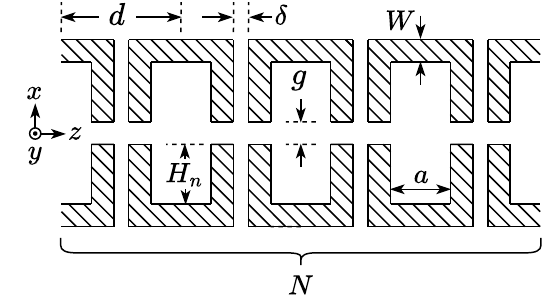}
    \caption{Proposed BPF structure based on SSPP with gapped internal unit cells. In the diagram, N = 4 as an example. Not to scale.}
    \label{fig:BPF_Internal_Structure}
\end{figure}

\begin{figure*}
    \centering
    \includegraphics[width=\linewidth]{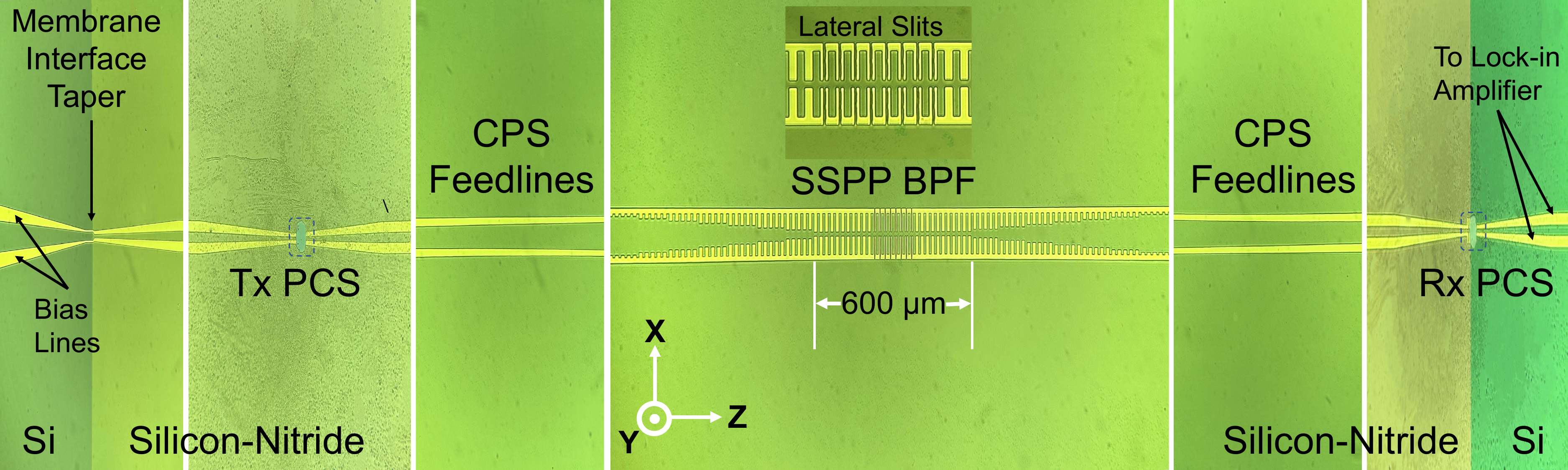}
    \caption{Fabricated SSPP-based BPF with gapped unit cells on a thin Si-N membrane excited by CPS feedlines. Key dimensions: Period $d$ = 20 \textmu m, SSPP groove width $a$ = 10 \textmu m, unit cell gap width $\delta$ = 3 \textmu m,  SSPP horizontal gap $g$ = 10 \textmu m (separation between top and bottom stubs), and groove depth $H_n$ = 40 \textmu m.}
    \label{fig:BPF_Internal_Fabricated}
\end{figure*}

Our previous work demonstrated a THz BPF that combines the effects of high-pass filtering via a capacitive TC and low-pass filtering by the SSPP waveguide \cite{haghighat2026terahertz}. In \cite{haghighat2026terahertz}, we found that the BPF has good performance and can operate in the THz region, but its design procedure can be cumbersome because both the upper and lower cut-off frequencies cannot be predicted by unit cell eigenfrequency simulations. The filter presented in this work can be initially designed using only eigenfrequency simulations of the unit cell, which expedites the design procedure. The SSPP-based THz BPF introduced in this paper is based on a \textit{gapped} internal CPS-SSPP unit cell. Illustrations of gapped and ungapped unit cells are shown in Table \ref{tab:unit_cells}. The addition of the gap converts the response from a LPF to a BPF. In this paper, we will investigate the complete BPF structure which means that we must first excite CPS feedlines (quasi-TEM) and then cascade a TC to convert the quasi-TEM mode to the SSPP TM mode \cite{unutmaz_investigation_2022, Haghighat2024_NSRep}. The TC consists of gradually lengthened SSPP stubs until the final maximum depth of corrugation is reached. For a substrate, we use a thin 1 \textmu m silicon nitride (SiN) membrane, which has been shown to reduce loss and dispersion at THz frequencies \cite{Levi_Smith_CPS_on_Si3N4_1st,Haghighat2024_NSRep}. Figure \ref{fig:BPF_Internal_Structure} shows the BPF, and Fig. \ref{fig:BPF_Internal_Fabricated} shows the fabricated SSPP-based BPF on the SiN membrane, along with photoconductive switches (PCS) for the generation and detection of THz signals.

\begin{table}[h]
\renewcommand{\arraystretch}{1.5}
\caption{Unit cell illustrations}
\label{tab:unit_cells}
\centering
\begin{tabular}{|c|c|}
\hline
Gapped Internal SSPP (this work) & Ungapped Internal SSPP  \cite{guo2018spoof_21} \\ \hline
   \includegraphics[width = 3cm]{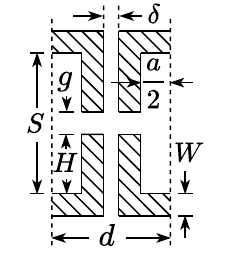}
   & \includegraphics[width = 3cm]{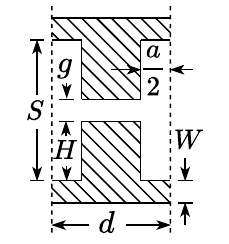} \\ \hline
   Bandpass Filter & Lowpass Filter \\ \hline
\end{tabular}
\end{table}

\begin{table*}[h!]
\centering
\caption{Comparison of Experimental Results for SSPP-based BPFs}
\label{tab:BPF_compare}
\begin{tabular}{m{3cm} m{2cm} m{2cm} m{2cm} m{2cm}}
\hline
Reference & $f_0$ (GHz) & FBW  & Min. IL (dB) \\ \hline \hline

\cite{guo_novel_BPF_IEEE_access_2018} & 5 & 120\% & 0.37 \\ \hline

\cite{Wang2019_BPF_SSPP_10GHz_IEEEAccess} & 10 & 36\% & 1.60 \\ \hline

 \cite{Wei2020_BPF_SSPP_IEEE_plasma_sci} & 3 & 43\% & 1.00 \\ \hline

\cite{Liu-Xu2022_BPF_SSPP_80GHz_TMTT} & 85 & 45\% & 0.58 \\ \hline

\cite{Feng-Xu2024_BPF_SSPP_30GHz} & 33 & 18\% & 0.50 \\ \hline

\cite{ren2020leaky} & 10 & 6\% & 3 \\ \hline

\cite{haghighat2026terahertz} & 1000 & 30\% & 7 \\ \hline

This Work  & 1000 & 25\% * & 7 ** \\ \hline \\
\parbox[t]{300pt}{* This FBW is related to the fabricated and measured structure, but it can be changed with geometry modifications.}
\parbox[t]{300pt}{** Simulated values considering all loss sources and with $\delta$ = 3 \textmu m and $N$ = 8.}

\end{tabular}
\end{table*}

Table \ref{tab:BPF_compare} provides a comparison of several SSPP-based BPFs with experimental results and our work. We were unable to find experimental results for many other SSPPs-based BPFs that were verified at THz frequencies; therefore, in Table \ref{tab:BPF_compare}, our structure clearly exhibits the highest insertion loss (IL) compared to its microwave counterparts that operate at less than 1/10th of the frequency. Also, the insertion loss is comparable to our previous design at 1 THz \cite{haghighat2026terahertz}. This increased insertion loss is primarily due to increased conduction losses at THz frequencies. We will briefly address the high insertion loss by comparing the SSPP-based BPF with the SSPP-based LPFs found in the literature. In \cite{unutmaz2025complete}, they measure the insertion loss to be between 5-10 dB at 0.275 THz. In \cite{xu2025terahertz}, they measure a 4 dB insertion loss at 0.14 THz. Therefore, we note that our proposed filter has an insertion loss comparable to other SSPP filters and operates at THz frequencies. We clarify that the goal of this research is not to propose a high performance optimized filter; this paper aims to demonstrate that the proposed proof-of-concept filter exhibits bandpass behavior at THz frequencies and to provide the reader with design guidance. Future work can improve the insertion loss by modifying the geometry. For example, simply increasing the conductor thickness will improve the insertion loss.

\section{Simulations and Design}
\label{sec:design}

\subsection{Simulation details}
\label{sec:simulation}

In the following sections, eigenfrequency and frequency domain (FD) simulations are performed using COMSOL version 6.4 for design and analysis purposes. The eigenfrequency simulations were performed on the SSPP unit cell. The FD simulations were performed on the complete structure. In all cases the following parameters are used: \(d = 2a = 20\) \textmu m, \(W = 10\) \textmu m, and $g$ = 10 \textmu m. The thickness of the SiN membrane is 1 \textmu m, relative permittivity \(\varepsilon_r = 7.6\), conductivity \(\sigma_{\text{Si}\text{N}} = 0\), and loss tangent \(\tan \delta_e = 0.00526\) \cite{Cataldo_Silicon_nitride_properties_2012}. The 200 nm thick conductors were modeled as a perfect electric conductor (PEC) or as gold using the Drude model with  $\varepsilon = 1 - \omega_p^2(\omega^2+j\omega/\tau)^{-1}$, where: $\omega_p = 2\pi$(2.2 PHz) and $\tau$ = 18 fs \cite{ordal1987optical}. Modeling the conductors as PEC or gold (Drude) was selected depending on the aim of the simulation, and the material is noted in the figure captions when relevant.

\subsection{Theory and dispersion analysis}
\label{sec:theory_dispersion}

In this section, we present a design procedure for the proposed SSPP BPF, which is based on the simulated dispersion analysis of the unit cell that forms the core of the filter. The initial geometric parameters for the unit cell are selected using design criteria found in the literature \cite{guo_spoof_2018, unutmaz_terahertz_2019,Haghighat_Internal_SSPP_24_OE}. The unit cell period, $d$, should be much smaller than the wavelength, $d \ll \lambda_0$ \cite{garcia-vidal_surfaces_2005,Haghighat2024_NSRep}. In this work, we operate near 1 THz, so $d \ll $ 300 \textmu m. In addition, $d$ should be large enough to ensure that the entire unit cell has sufficient resolution using photolithography fabrication; our photolithographic resolution is 3 \textmu m. In this work, we select $d$ = 20 \textmu m. Next, the aperture, $a$, can be selected between $0.1d< a < 0.9d$, but selecting $a = 0.5d = 10$ \textmu m results in low loss and good performance. The width of the gaps, $\delta$, primarily controls the lower cutoff frequency. The details of this parameter are discussed later, but we select $\delta$ = 3 \textmu m, which corresponds to our limitation in feature size via photolithography. The gap between the internal SSPP stubs, $g$, does not have a large impact on the filter performance, but it is important that $g$ is equal to or greater than the conductor spacing at the Tx PCS (see Fig. \ref{fig:BPF_Internal_Fabricated}) to ensure that dielectric breakdown does not occur when applying a bias voltage. Thus, we select $g$ = 10 \textmu m to satisfy this criterion, but larger values could also have been chosen if desired.

\begin{figure}[H]
    \centering
    \includegraphics[width=0.9\linewidth]{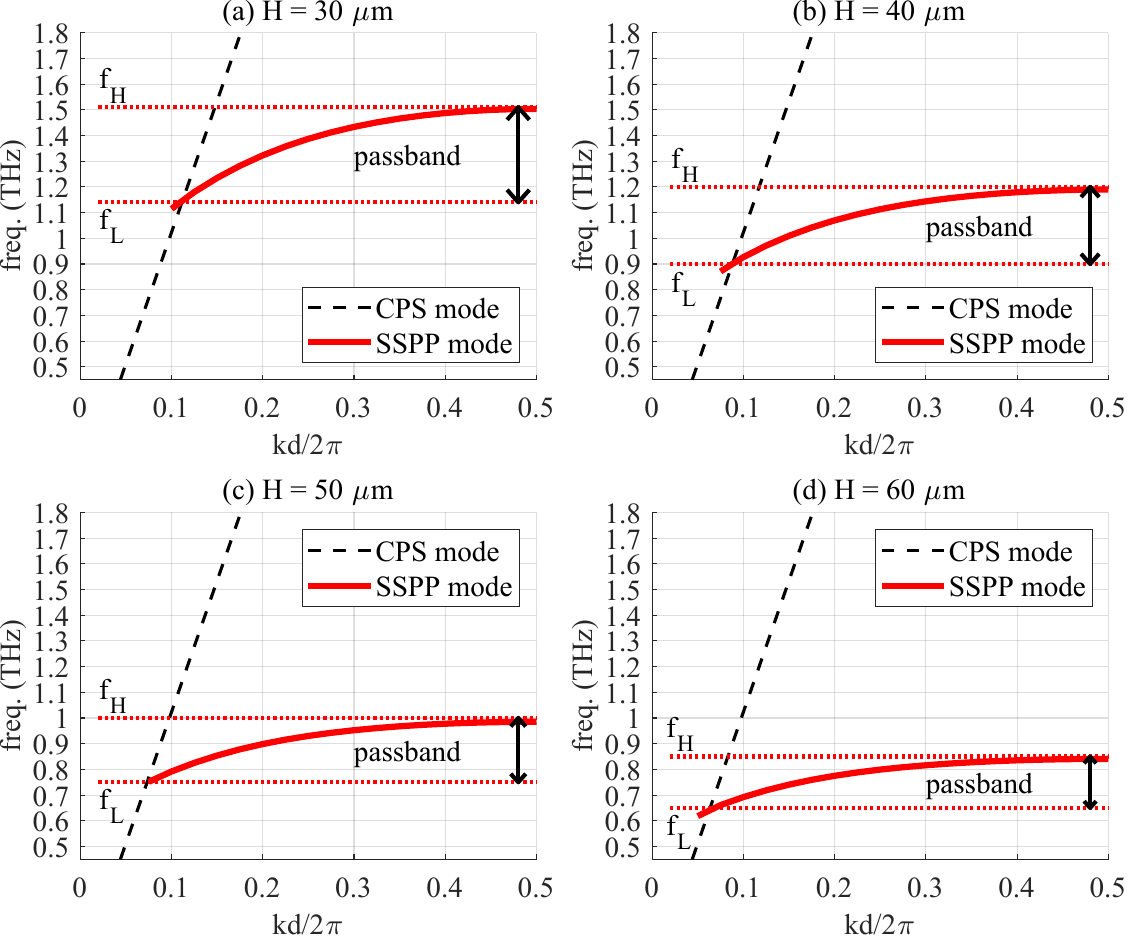}
    \caption{Dispersion curves of gapped SSPP unit cell for $H_n$ = 30, 40, 50, and 60 \textmu m showing bandpass filtering behavior. Conductors are modeled as PECs.}
    \label{fig:BPF_Internal_Dispersion_Curves}
\end{figure}

\begin{figure}[H]
    \centering
    \includegraphics[width=0.9\linewidth]{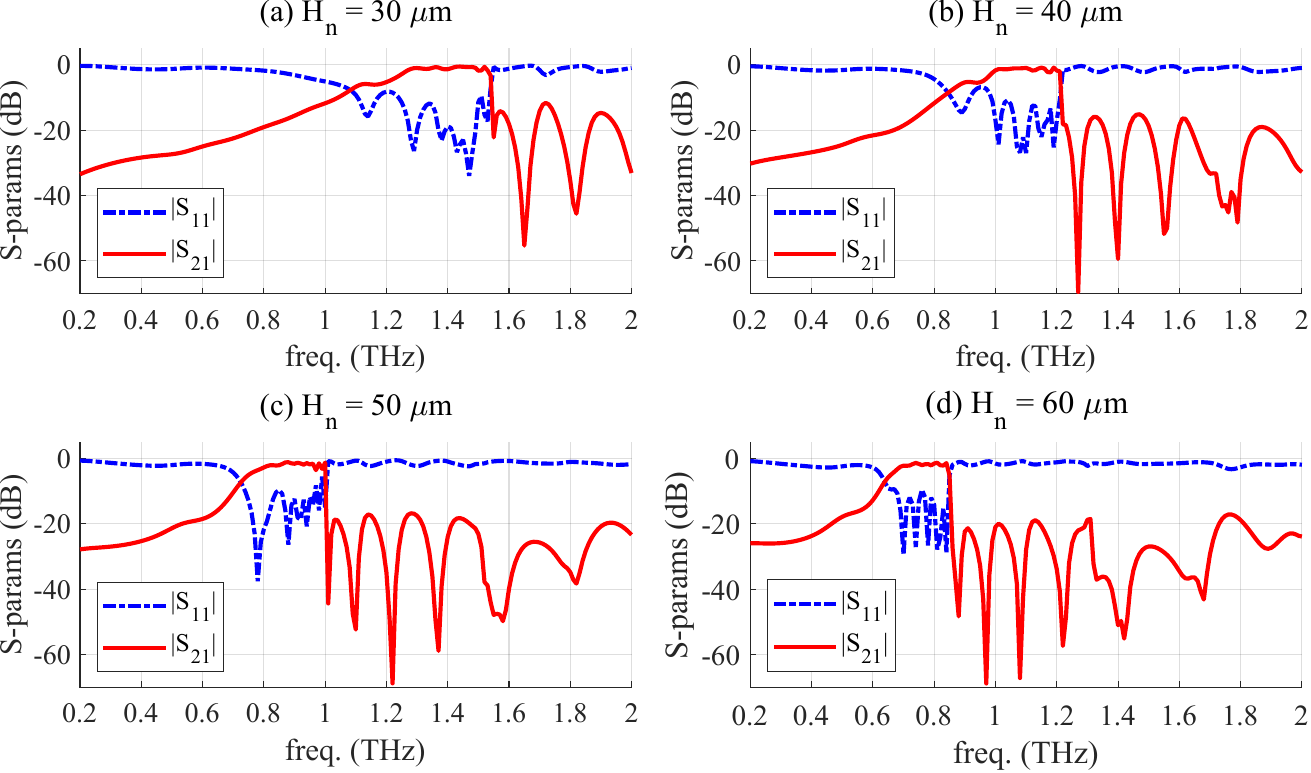}
    \caption{Simulated $S_{21}$ for the gapped SSPP BPF with $H_n$ = 30, 40, 50, 60 \textmu m. For this simulation $\delta$ = 3 \textmu m and N = 8. Conductors are modeled as PECs.}
    \label{fig:BPF_Internal_S21_30_40_50}
\end{figure}

The design procedure starts by using an eigenfrequency simulation of the unit cell to determine the cutoff frequencies. Figure \ref{fig:BPF_Internal_Dispersion_Curves}(a-d) illustrates the simulated dispersion diagram for $H_n$ = 30, 40, 50, and 60 \textmu m, respectively. The dispersion curves start at the lower cutoff point, $f_L$, and end at the higher band edge frequencies, $f_H$. $f_H$ is determined by the SSPP band-edge defined as the frequency where $kd = \pi$, which is mainly controlled by $H_n$ \cite{Haghighat2024_NSRep,Haghighat_Internal_SSPP_24_OE}. $f_L$ is obtained from where the dispersion curve for the CPS mode (similar to a light line with $\varepsilon_{r,eff} = 2.0$ \cite{haghighat2026terahertz}) intersects the SSPP mode. These frequencies align with the cut-off frequencies observed from the S-parameters obtained from full-wave simulations that are shown in Fig. \ref{fig:BPF_Internal_S21_30_40_50}. We clarify that in Fig. \ref{fig:BPF_Internal_S21_30_40_50} we simulate: Port 1 $\rightarrow$ CPS feedlines $\rightarrow$ TC $\rightarrow$ SSPP unit cells $\rightarrow$ TC $\rightarrow$ CPS feedlines $\rightarrow$ Port 2. This is required to ensure that the SSPP mode is excited. Also, for this simulation, the conductors were modeled as PECs to clearly illustrate the cutoff frequencies. Both Figs. \ref{fig:BPF_Internal_Dispersion_Curves} and \ref{fig:BPF_Internal_S21_30_40_50} show that changing $H_n$ changes both $f_L$ and $f_H$.

\begin{figure}[H]
    \centering
    \includegraphics[width=0.7\linewidth]{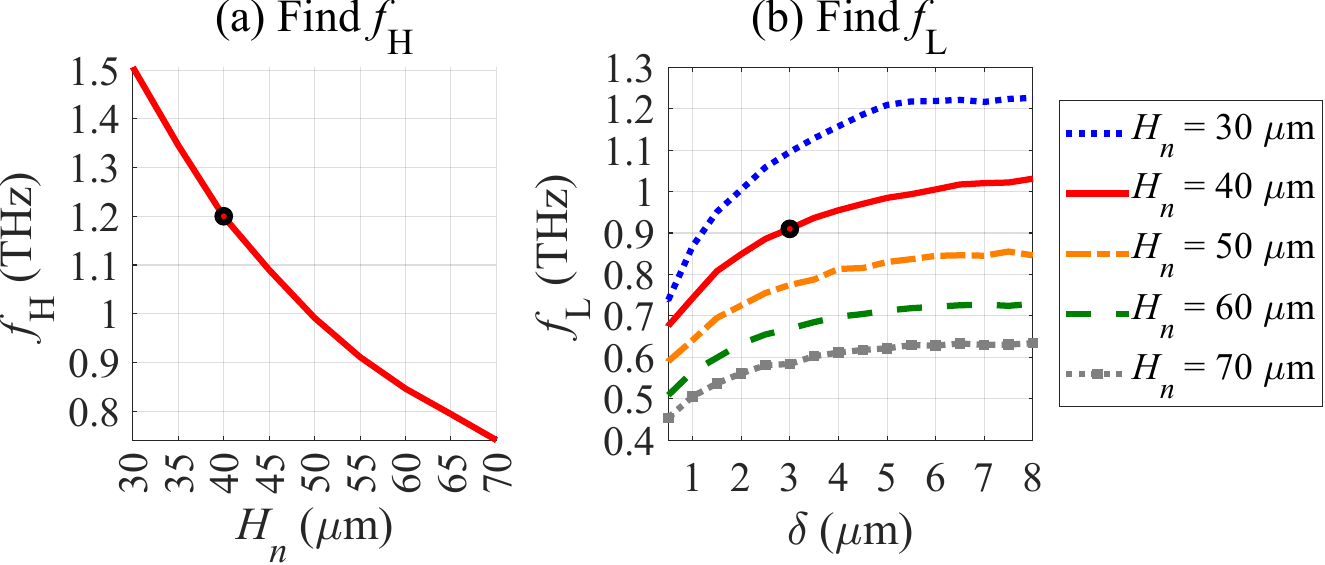}
    \caption{Unit cell design for higher and lower cutoff frequencies. For these curves: $d = 2a$ = 20 \textmu m, $W$ = 10 \textmu m, and g = 10 \textmu m. Conductors are modeled as PECs. (a) Determining the required $H_n$ to achieve a specified $f_H$. (b) Determining the required $\delta$ to achieve a specified $f_L$.}
    \label{fig:Find_fL_and_fH}
\end{figure}

\begin{figure}[H]
    \centering
    \includegraphics[width=0.5\linewidth]{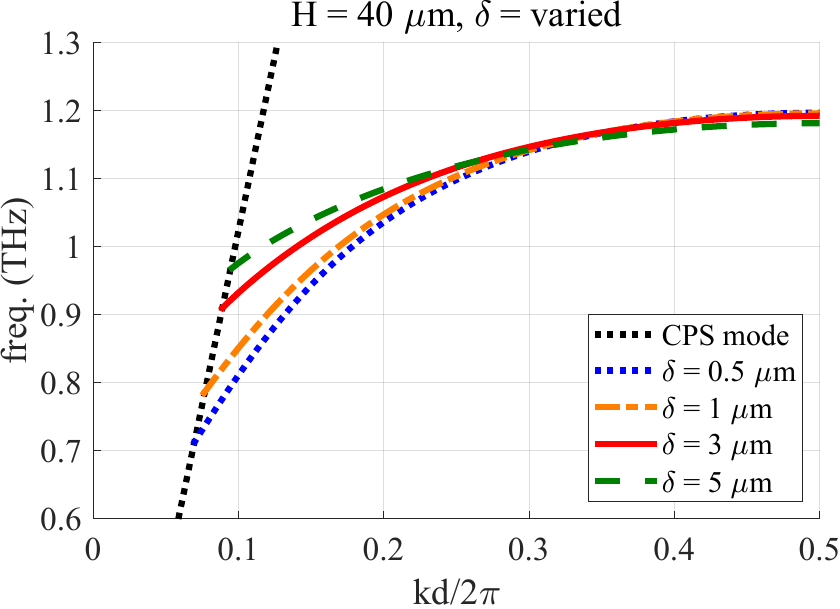}
    \caption{Simulated dispersion curves for different $\delta$ parameters to design the lower cut-off of the proposed BFP without affecting the higher cut-off for the  structure with \mbox{$H_n$ = 40 \textmu m.} Conductors are modeled as PECs.}
    \label{fig:BPF_Internal_dispersion_delta_sweep}
\end{figure}

\begin{figure}
    \centering
    \includegraphics[width=0.9\linewidth]{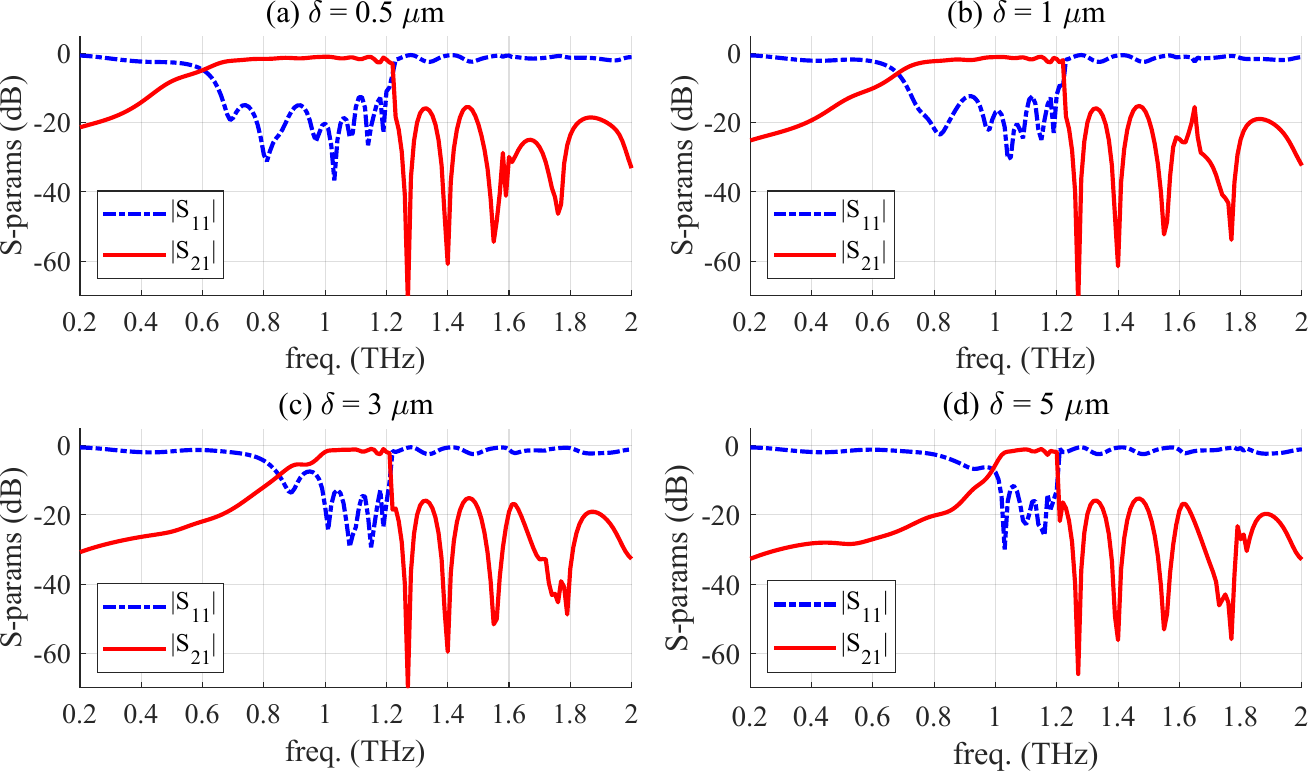}
    \caption{Simulated S-parameters for different $\delta$ which illustrates the change in the lower cut-off frequency of the BPF without affecting the higher cut-off with $H_n$ = 40 \textmu m. In all cases the conductors are modeled as PECs and the passband insertion loss is approximately 1 dB.}
    \label{fig:BPF_Internal_S21_delta_sweep_lower_cutoff}
\end{figure}

The high cut-off frequency, $f_H$, is primarily controlled by $H_n$ \cite{garcia-vidal_surfaces_2005,Haghighat2024_NSRep}. The relationship between these variables is shown in Fig. \ref{fig:Find_fL_and_fH}a, which was constructed from an eigenfrequency simulation where $W$ = 10 \textmu m, $d = 2a $ = 20 \textmu m, and $g$ = 10 \textmu m. The black dots on Fig. \ref{fig:Find_fL_and_fH} correspond to our experimental structure. For this proof-of-concept filter, we select $H_n$ = 40 \textmu m, which corresponds to a $f_H$ of 1.2 THz. The lower cut-off frequency, $f_L$, is controlled by varying $H_n$ and $\delta$. However, we found that $\delta$ can independently adjust $f_L$ without significantly impacting $f_H$. Therefore, the bandwidth (BW) of the filter can be adjusted by changing $\delta$, and consequently varying $f_L$ as shown in Fig. \ref{fig:BPF_Internal_dispersion_delta_sweep}. The relationship between $\delta$, $H_n$, and $f_L$ is plotted in Fig. \ref{fig:Find_fL_and_fH}b, which was constructed from an eigenfrequency simulation where $W$ = 10 \textmu m, $d = 2a$ = 20 \textmu m, and $g$ = 10 \textmu m. In summary, to design a gapped SSPP unit cell BPF, first use Fig. \ref{fig:Find_fL_and_fH}a to find the $H_n$ that corresponds to the desired $f_H$. Next, use Fig. \ref{fig:Find_fL_and_fH}b to find the $\delta$ that corresponds to the desired $f_L$. Figure \ref{fig:BPF_Internal_S21_delta_sweep_lower_cutoff}(a-d) illustrates the simulated S-parameters for $\delta$ = 0.5, 1, 3, and 5 \textmu m, respectively, with $H_n$ = 40 \textmu m. In Fig.\ref{fig:BPF_Internal_S21_delta_sweep_lower_cutoff} it is observed that $\delta$ can be used to independently control $f_L$ of the BPF without significantly affecting $f_H$. The results presented in Fig. \ref{fig:BPF_Internal_S21_delta_sweep_lower_cutoff} are in good agreement with the dispersion analysis shown in Fig. \ref{fig:BPF_Internal_dispersion_delta_sweep}.

\begin{figure}[H]
    \centering
    \includegraphics[width=0.9\linewidth]{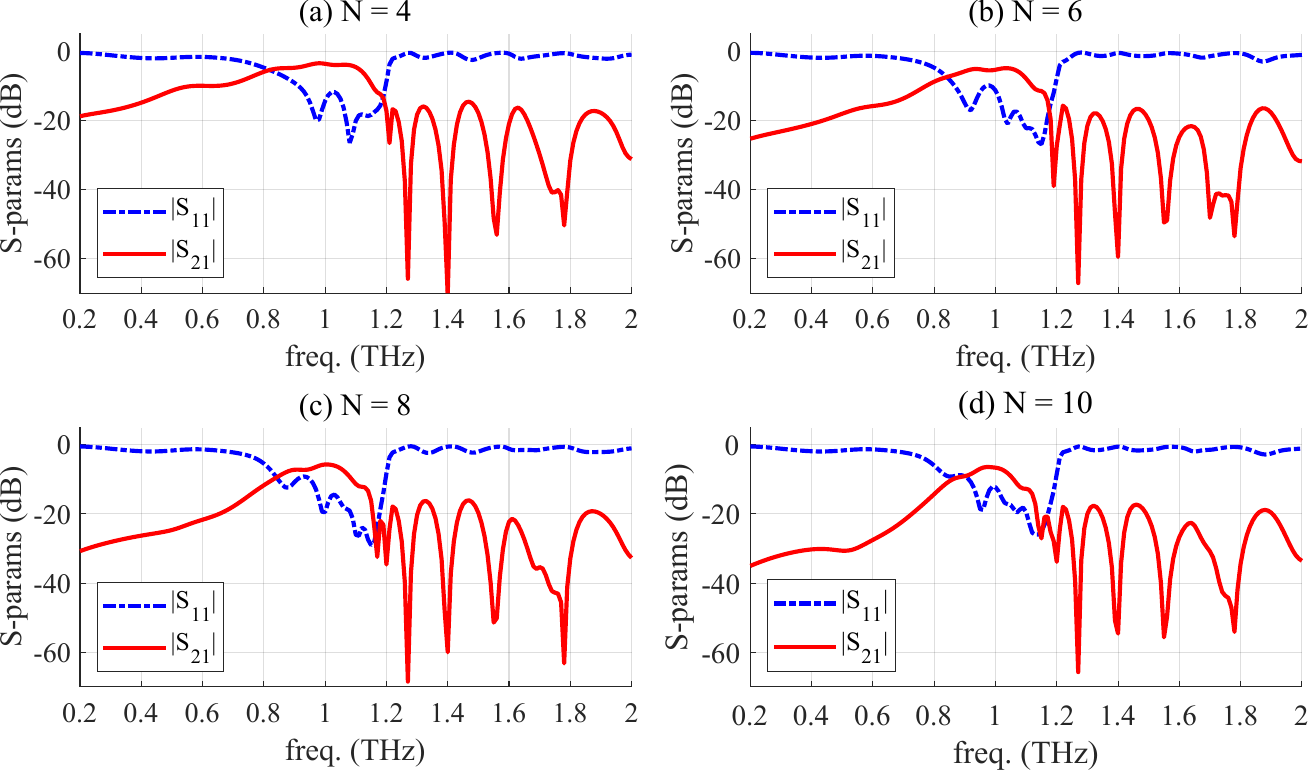}
    \caption{Simulated S-parameters for different $N$ which illustrates the change in the lower cut-off frequency roll-off rate and insertion loss with $H_n$ = 40 \textmu m. In all cases the conductors are modeled as gold (Drude). When N = 4 the passband IL is -4 dB. When N = 10 the IL is -8 dB.}
    \label{fig:BPF_Internal_S21_H40_Au200nm_N_sweep}
\end{figure}

Upon completion of the design of the unit cells, the number of cascaded unit cells, $N$, must be determined. Figure \ref{fig:BPF_Internal_S21_H40_Au200nm_N_sweep} plots the S-parameters for N = 4, 6, 8, and 10. In all of these simulations, the conductors were modeled as gold (the aforementioned Drude model). When selecting $N$, there is a trade-off between the insertion loss and the roll-off rates. As $N$ increases, the roll-off rates will increase; however, the passband insertion loss also increases, as illustrated in Fig. \ref{fig:BPF_Internal_S21_H40_Au200nm_N_sweep}. The lower roll-off rate is dependent on the number of periods, $N$, which should be selected based on the requirements for the specific application. For our proof-of-concept design, we selected $N$ = 8 to achieve acceptable roll-off rates and insertion loss ($\approx$ 7 dB). Using these parameters, we can safely anticipate that our experiment will be able to resolve the transmitted signal \cite{haghighat2026terahertz}. In Fig. \ref{fig:BPF_Internal_S21_H40_Au200nm_N_sweep}, there is a rounding of $|S_{21}|^2$ near $f_H$ when conductor loss is included. This is expected because the group velocity trends towards zero at the band edge, and thus frequencies near the band edge experience greater interaction with the conductor and therefore exhibit increased attenuation.

An approach to reducing passband insertion loss, with a fixed $N$, is to decrease $\delta$. Note, this also affects the lower cutoff frequency, which must be considered during design iterations. We selected $\delta$ = 3 \textmu m, which corresponds to the minimum feature size available in our fabrication process. We note that it is possible to achieve $\delta$ $\leq$ 3 \textmu m and reduce the insertion loss with more precise lithography and fabrication methods. Also the conductor thickness can be increased to reduce conductor loss. Lastly, increasing the width of the CPS lines, $W$, and the separation between the CPS lines, $S$, can help reduce the insertion loss, but it can result in higher radiation losses depending on the application.

\begin{figure*}[t]
    \centering
    \includegraphics[width=0.8\linewidth]{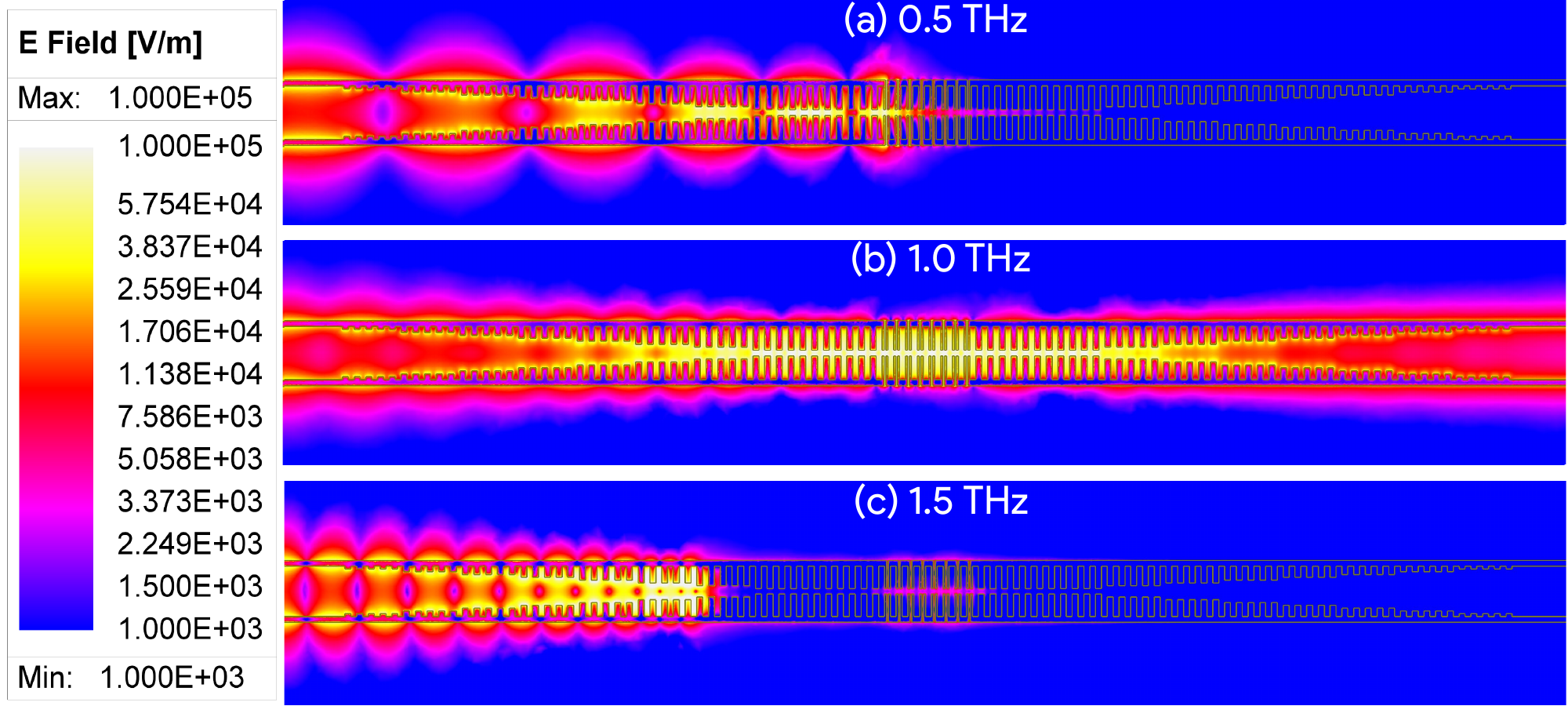}
    \caption{Electric field plots of the proposed THz SSPP band-pass filter with $H_n$ = 40 \textmu m at 0.5 THz (lower stopband), 1 THz (passband), and 1.5 THz (higher stopband), illustrating passband transmission and out-of-band rejection.}
    \label{fig:Filed_Plots_BPF_Internal}
\end{figure*}

To visualize the filtering operation, electric field plots for the BPF with \(H_n = 40\) \textmu m are plotted in Fig. \ref{fig:Filed_Plots_BPF_Internal} at 1 THz (in the passband), 0.5 THz, and 1.5 THz (the lower and higher stopbands, respectively). These plots also illustrate the strong field confinement capability of the proposed SSPP structure inside the SSPP unit cell gaps and grooves.

We briefly comment on the transition circuit. The core of the structure with fixed $H_n$ should not be directly connected to the CPS feedlines because there is a large mode mismatch between the CPS (quasi-TEM) and the SSPP (quasi-TM). Therefore, a TC with a gradual increase in the depth of the SSPP grooves is required \cite{Haghighat2024_NSRep}. In this work, we used a TC with stepped linear growth up to $H_n$ for the fabricated device, where each TC stub with a height of $H_i$ ($ 1\leq i \leq 7$) is repeated five times and $H_i$ increases linearly in each section (see Fig. \ref{fig:BPF_Internal_Structure}) to reach the maximum depth of SSPP (i.e., $H_n$). The reason for choosing this TC configuration is its ability to provide more rejection at stopband frequencies through the repetition of stubs, as each stub length corresponds to the rejection of a single frequency. We note that our selected method is not a strict requirement, and the selection of transition circuits is somewhat flexible, provided there is a gradual introduction of stub lengths. 

\noindent The design procedure is summarized in the following steps:
\begin{enumerate}
    \item Specify the filter requirements: $f_H$, $f_L$, and IL. We selected $f_H$ = 1.2 THz, $f_L$ = 0.9 THz, and IL = 7 dB.
    \item Determine $H_n$ for the specified $f_H$. The relationship between $H_n$ and $f_H$ is determined by an eigenfrequency simulation. We provide Fig. \ref{fig:Find_fL_and_fH}a which was generated when $\delta$ = 3 \textmu m. Note that $\delta$ = 3 \textmu m can be used as an initial value for most designs (see Fig. \ref{fig:BPF_Internal_dispersion_delta_sweep} when $kd = \pi$). However, $\delta$ will be modified in the next step.
    \item Determine $\delta$ for the specified $f_L$. This can be performed using an eigenfrequency simulation. 
    \item Determine $N$. Perform a frequency domain simulation on the complete filter to determine the number of gapped unit cells, $N$. The resultant S-parameters quantify passband insertion loss and filter performance. Figure \ref{fig:BPF_Internal_S21_H40_Au200nm_N_sweep} illustrates the effect of $N$ on the S-parameters, where it is shown that with $N$ = 8, we achieve good roll-off above and below the passband.
\end{enumerate}

\section{ Experimental Setup and Fabrication of PCS}
\label{sec:exp}

\begin{figure}
    \centering
    \includegraphics[width=0.5\linewidth]{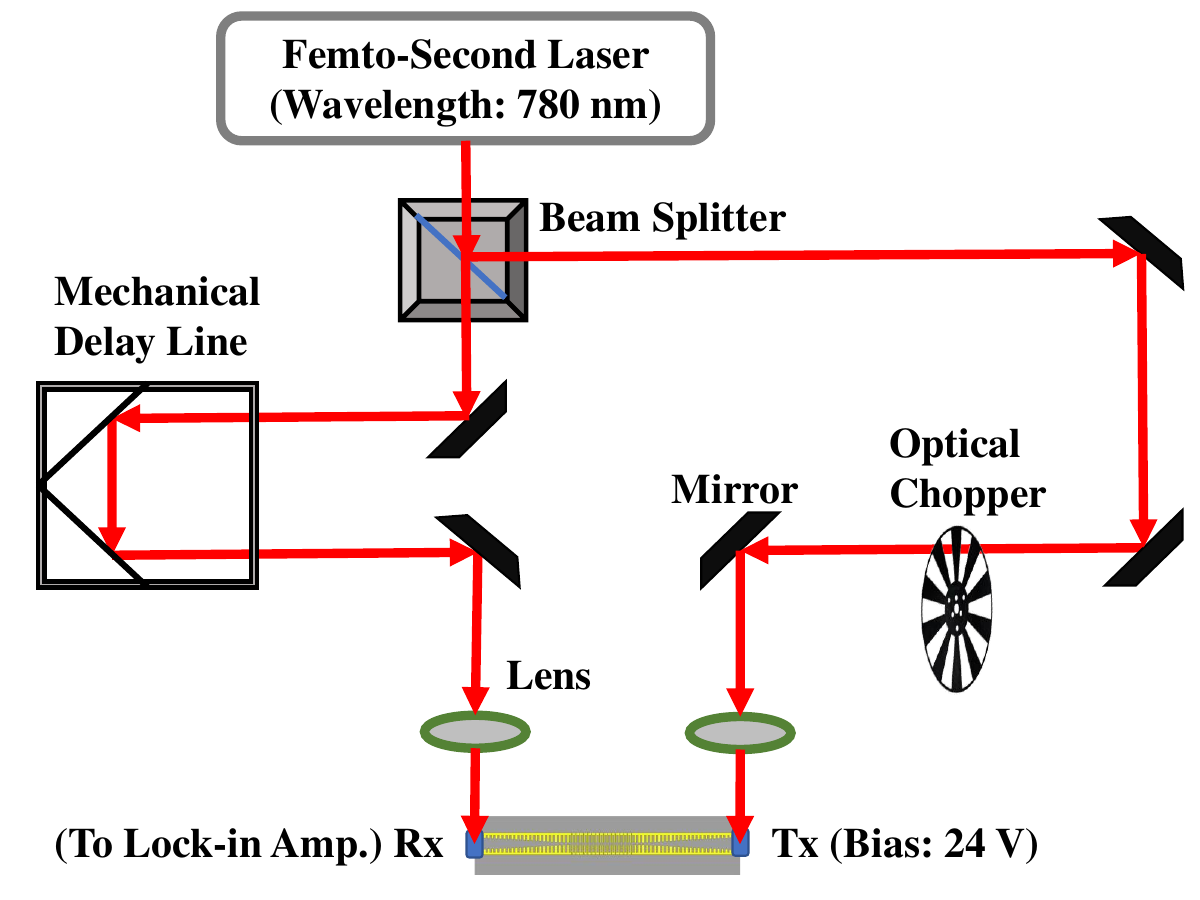}
    \caption{Measurement Setup}
    \label{fig:BPF_Internal_Measurement_Setup}
\end{figure}

The experimental setup used for the measurements is based on a modified THz time-domain spectroscopy (THz-TDS) system, as outlined in \cite{Levi_Smith_CPS_on_Si3N4_1st}, and illustrated in Fig. \ref{fig:BPF_Internal_Measurement_Setup}. This system features a femtosecond  pulsed laser operating at a wavelength of 780 nm, with a pulse width of 90 fs, a repetition rate of 80 MHz, and an average output power of 27 mW. The laser beam is focused onto photoconductive switches (PCS) located at the transmitter (Tx) and receiver (Rx) positions on the SiN membrane (see Fig. \ref{fig:BPF_Internal_Fabricated}). These switches function as sources and detectors of THz signals through the processes of photoconductive switching and sampling \cite{Levi_Smith_CPS_on_Si3N4_1st}.

The PCSs measure 70 \textmu m $\times$ 40 \textmu m $\times$ 1.5 \textmu m, with a gap of 5 \textmu m between the gold contacts. They are bonded using water droplets through Van der Waals (VDW) forces, as detailed in \cite{VDW1990}. The transmitted signal is reconstructed by translating the mechanical delay line and measuring the receiver current with a lock-in amplifier, following the methods used in standard THz-TDS setups \cite{Levi_Smith_CPS_on_Si3N4_1st,Haghighat2024_NSRep}.

The fabrication process for the photoconductive switches (PCSs) involves several steps. First, a low-temperature grown gallium arsenide (LT-GaAs) layer is deposited on a sacrificial aluminum arsenide (AlAs) layer, which rests on a semi-insulating GaAs substrate. Next, photolithography is applied to the LT-GaAs surface to create gold (Au) contacts. Each PCS region is masked and subjected to wet etching with citric acid and hydrogen peroxide to define the PCS thickness. After cleaning and re-masking with etch-resist wax, the LT-GaAs layer is dissolved in hydrofluoric acid (HF), separating it from the AlAs layer. Further steps are taken to detach the remaining LT-GaAs film, resulting in multiple active regions of LT-GaAs PCSs \cite{Rios2015_bowtie_PCA}.

\section{Experimental Results and Discussion}

\begin{figure*}
    \centering
    \includegraphics[width=0.9\linewidth]{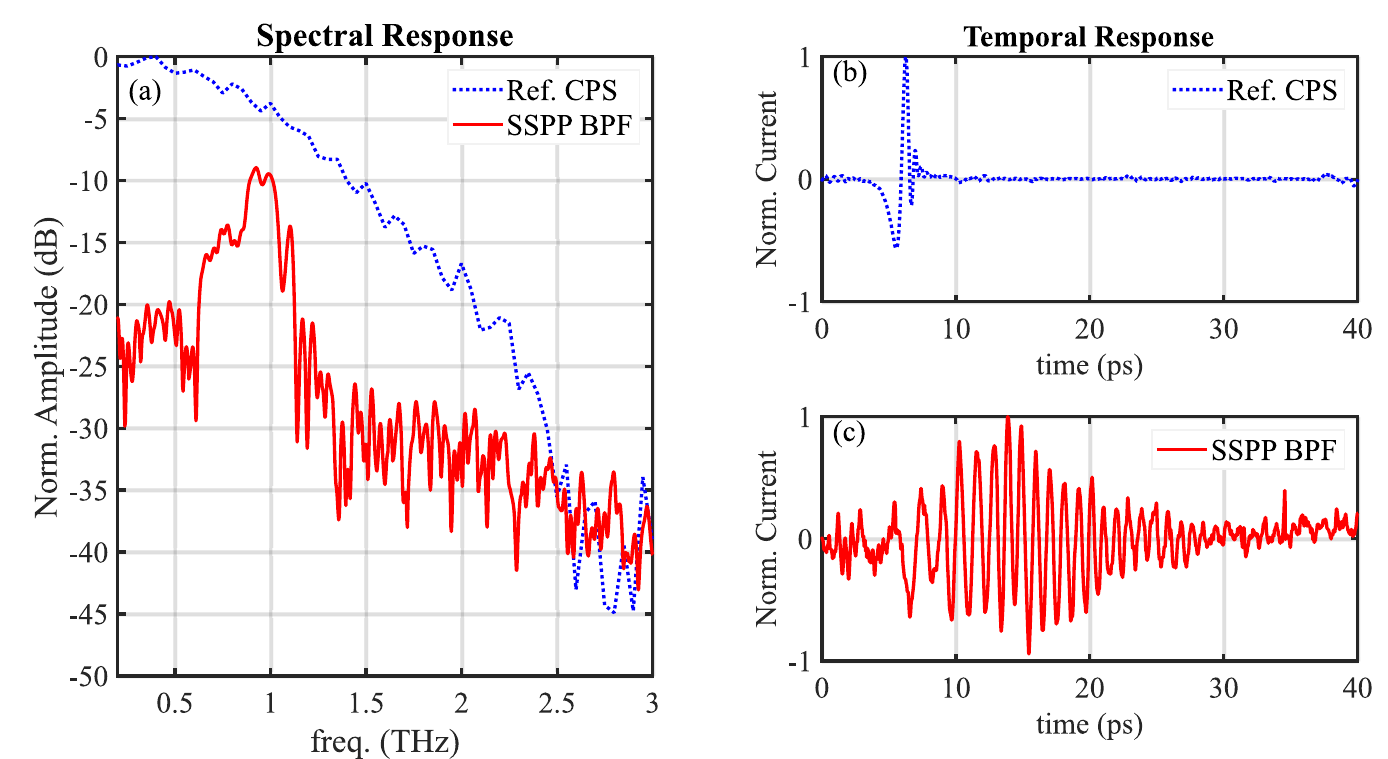}
    \caption{Measurement results of the proposed BPF with $H_n$ = 40 \textmu m versus a reference CPS with the same length}
    \label{fig:BPF_Internal_Measurement_Results}
\end{figure*}

The experimental results of the THz BPF that uses gapped unit cells with $H_n$ = 40 \textmu m are presented in Fig. \ref{fig:BPF_Internal_Measurement_Results}, showing both the temporal responses of the BPF and a reference CPS (transient output pulses). A reference device is a CPS feedline with the same length as the SSPP BPF. The temporal response of the reference exhibits an anticipated broadband pulse with a single peak, while the BPF response manifests as a narrowband pulse containing multiple oscillations. Additionally, Fig. \ref{fig:BPF_Internal_Measurement_Results} illustrates the corresponding spectral responses on a single subplot (left side) derived from applying the discrete Fourier transform (DFT) to the temporal responses. The spectral response of the BPF exhibits an abrupt roll-off at the expected cut-off frequencies of 0.9 THz (simulation predicted 0.91 THz) and 1.15 THz (simulation predicted 1.2 THz). The higher band-edge frequency aligns with the SSPP band-edge as predicted by dispersion curves (Fig. \ref{fig:BPF_Internal_Dispersion_Curves}b) and S-parameter simulations (Fig. \ref{fig:BPF_Internal_S21_H40_Au200nm_N_sweep}c).

Going forward, we believe that the proposed gapped unit cell SSPP BPF will be a useful addition to the THz community due to the straightforward design procedure that we presented and its performance characteristics at THz frequencies. While we recognize that a 7 dB insertion loss is non-ideal, we reiterate that there are few planar/non-planar alternatives (with experimental results at THz frequencies). Also, the proposed integrated filter has minimal coupling loss since the PCS's are directly connected to the transmission line, which mitigates the broadband coupling loss that other technologies must manage properly.

A couple of comments on interpreting the experimental results. First, it is important to address why the experimental spectral responses exhibit amplitude decay with respect to increasing frequency. This occurs because the transient signal has a finite duration, which implies that the spectral response must decay with increasing frequency. This means that the dominant frequency dependent `loss' for the spectral response traces in Fig. \ref{fig:BPF_Internal_Measurement_Results} is not truly `loss'; it simply indicates that the generated and detected signal is frequency dependent. Regardless, there is sufficient information contained within the received signal to demonstrate that signal filtering is occurring, which is in agreement with the filter design. Second, when investigating BPF's, the detected signal strength is reduced compared to other configurations (i.e., the reference, low-pass, high-pass, etc.) because the majority of the source signals spectral content is removed by the BPF by the time the signal reaches the receiver. While generally this is advantageous and is the purpose of a BPF, it presents a minor adverse artifact during the experiment because the signal-to-noise ratio decreases. This is observed when comparing the temporal responses for the signals in Fig. \ref{fig:BPF_Internal_Measurement_Results} since the reference signal has less noise than the filtered signal. This is a challenge that is observed during experimentation and can make system alignment more difficult.


\section{Conclusion}

In this work, we introduce a novel terahertz (THz) band-pass filter (BPF) based on spoof surface plasmon polariton (SSPP) with gapped unit cells, developed from coplanar strips (CPS) with internal stubs. The filter achieves high-frequency rejection through low-pass SSPP characteristics and effectively blocks low frequencies through gaps within the structure. By adjusting the geometrical parameters, particularly the $\delta$ and $H_n$ parameters, the higher and lower cut-off frequencies and the bandwidth of the filter can be customized, providing flexibility in design. Experimental validation of the fabricated BPF, with a center frequency of approximately 1 THz, demonstrates promising performance, with measured passband transmission closely matching the simulations regarding cut-off frequencies. This validates the effectiveness of the proposed SSPP structure in achieving the desired bandpass filtering functionality. This work highlights the potential of SSPP-based filters in advanced THz applications, such as sensing, with the ability to tailor frequency responses through simple geometrical modifications. To our knowledge, this is the first experimental demonstration of a gapped SSPP filter at THz frequencies.

\section*{Acknowledgments}

This work was supported by an NSERC Discovery Grant. The authors thank 4D LABS at Simon Fraser University for the fabrication of the waveguides and the thin membrane, and also the Center for Advanced Materials and Related Technology (CAMTEC) at the University of Victoria for providing Nanofab facilities for the fabrication of the PCS devices.

\vspace{5pt}

\noindent \textbf{Funding} This work was supported by the Natural Sciences and Engineering Research Council of Canada (NSERC) (RGPIN-2022-03277).

\noindent \textbf{Availability of data and materials} : Data supporting the findings of this study are available from the corresponding author, L.S., upon reasonable request.

\section*{Declarations}

\noindent \textbf{Ethical Approval} Not applicable.

\vspace{8pt}

\noindent \textbf{Competing Interests} The authors declare no competing interests.

\bibliographystyle{unsrt}
\bibliography{SSPP_references}

\end{document}